\documentclass[aps,prd,preprint,nofootinbib,showpacs]{revtex4}

\def\lb{\lbrack}
\def\rb{\rbrack}

 \def\Slash#1{
  \begin{picture}(5,6)(0,0)
  \put(-.7,-1.2){\line(5,6)6}
  \end{picture}
  \kern-.8em#1}
 \def\slash#1{
  \begin{picture}(5,6)(0,0)
  \put(-1.5,-1.7){\line(5,6)5}
  \end{picture}
  \kern-.8em#1}

\def\Sn{\Slash \nabla}
\def\sd{\Slash \partial}

\def\e{\epsilon}
\def\gg5{\gamma_5}
\def\hg5{\hat{\gamma}_5}
\def\g4{\gamma_4}

\def\Qlatmr1{Q_{lat}^{(m=r=1)}}

\def\be{\begin{eqnarray}}
\def\ee{\end{eqnarray}}

\def\bta{\bar{\eta}}
\def\bpsi{\bar{\psi}}
\def\gmu{\gamma_{\mu}}
\def\g5{\gamma_5}
\def\tg{\tilde{\gamma}}
\def\tW{\tilde{W}}

\begin{document}
 
\title{The rooting issue for a lattice fermion formulation similar to
staggered fermions but without taste mixing}

\author{David H. Adams}
\email{dadams@phya.snu.ac.kr}

\affiliation{Department of Physics and Astronomy, Seoul National University,
Seoul, 151-747, South Korea}

\date{March 14, 2008}

\begin{abstract}

To investigate the viability of the 4th root trick for the staggered fermion
determinant in a simpler setting, we consider a two taste (flavor) lattice
fermion formulation with no taste mixing but with exact taste-nonsinglet chiral 
symmetries analogous to the taste-nonsinglet $U(1)_A$ symmetry of staggered 
fermions. M.~Creutz's objections to the rooting trick apply just as much in this
setting. To counter them we show that the formulation has robust would-be
zero-modes in topologically nontrivial gauge backgrounds, and that these
manifest themselves in a viable way in the rooted fermion determinant
and also in the disconnected piece of the pseudoscalar meson propagator
as required to solve the U(1) problem. Also, our rooted theory is heuristically
seen to be in the right universality class for QCD if the same is true for an 
unrooted mixed fermion action theory.

\end{abstract}


\maketitle

\section{Introduction}

The use of dynamical staggered fermions in lattice QCD simulations has made it
possible to obtain results with unprecedented high precision 
\cite{Davies(PRL),MILC,Allison}.
However, this approach is controversial due to the use of the ``4th root 
trick'': A staggered lattice fermion corresponds to 4 continuum fermion flavors 
(nowadays called tastes to distinguish them from the actual quark flavors),
so the fermion determinant for each dynamical quark flavor is represented
by the 4th root of the corresponding staggered fermion determinant. Since this
formulation is not manifestly a local lattice field theory there is a danger
that it might not be in the right universality class for QCD. (In fact it has
been argued \cite{BGS} that this lattice theory is necessarily non-local but 
with locality being restored in the continuum limit 
\cite{Shamir}.\footnote{In the free field case (at least for $m\ne0$) the rooted 
staggered formulation corresponds to a local field theory already at 
non-vanishing lattice spacing \cite{DA(PRD)}, as was also confirmed numerically
\cite{Peardon}.})
Because of the high stakes, this has become a prominent, hotly debated issue
in the lattice community. E.g., it has been the topic of 5 plenary talks at the
last 4 annual lattice field theory conferences; the corresponding proceedings 
papers \cite{DA(04),Durr(05),Sharpe(06),Kronfeld(07),Creutz(07)} can be 
consulted for reviews from various perspectives. 

While the results to date are in excellent, unprecedented agreement
with experiment, a major question regarding the 4th root trick for staggered 
fermions is whether it can work in situations where chirality is important.
This includes in particular producing the large mass of the $\eta'$ meson where
existence of fermionic zero-modes with definite chirality and their connection 
with topological charge of lattice gauge fields via the Index Theorem plays an 
essential role \cite{thooft}.\footnote{Efforts to calculate the 
$\eta'$ mass to high precision with dynamical staggered fermions are currently 
underway \cite{Gregory1,Gregory2}. In the meantime,
encouraging evidence that this formulation is able to correctly reproduce 
topological aspects of QCD has been given in Ref.~\cite{Bernard(top)}
where results for the topological susceptibility were presented.} 
M.~Creutz has argued against this in a series of 
papers \cite{Creutz1,Creutz2,Creutz(chiral),Creutz(chiral)2nd,Creutz(thooft)} 
based on the fact that the the taste-nonsinglet $U(1)$ 
chiral symmetry of staggered fermions implies properties of the rooted staggered 
fermion determinant that do not hold for a genuine single-flavor fermion 
determinant. The subsequent rebuttals of these arguments 
\cite{BGSSrebut1,BGSSrebut2} rely
to a large extent on invoking full taste symmetry restoration on the continuum 
limit. However, Creutz challenges whether this can actually occur in a
way that correctly reproduces nonperturbative effects connected with chirality.
In this situation it is desirable to have a simpler setting where the same 
issues arise and where they can be studied more explicitly. We provide and
study such a setting in the present paper.   

The paper is organized as follows. In sect.~II we contrast a general mixed fermion
formulation with a rooted formulation based on a 2-taste lattice Dirac operator
without taste-mixing, showing heuristically that if the former is in the right 
universality class for QCD then so is the latter. In sect.~III we introduce
the specific 2-taste lattice Dirac operator with exact taste-nonsinglet chiral 
symmetries on which the rooted formulation studied in this paper is based.
In sect.~IV we study the properties of the single-flavor fermion formulations
based on the 1-taste lattice Dirac operators making up our 2-taste operator,
and use this to derive properties of the rooted formulation based on the latter.
In sect.~V we discuss the pseudo-scalar propagator in the rooted formulation,
and conclude with a discussion in sect.~VI. A relation between, and differences
between, our 2-taste formulation and the 2-flavor Wilson fermion theory with 
twisted mass is discussed in an appendix.

\section{Prelude: mixed fermion action versus a rooted formulation}

We begin with some general remarks on generating functionals for lattice
fermions (in a fixed gauge field background, in Euclidean spacetime). For a
single quark flavor described by a lattice Dirac operator $D_1$ the generating 
functional is
\be
Z_f(\eta,\bta)&=&\int d\psi d\bpsi\,
e^{-\bpsi D_1\psi+\bpsi\bta+\bta\psi} \nonumber \\
&=&\det D_1\,e^{\bta D_1^{-1}\eta}
\label{1}
\ee
For a ``mixed fermion action'' where the sea quark is described by $D_1$
and the valence quark by another lattice Dirac operator $D_2$ the generating
functional becomes
\be
Z_{f,mixed}(\eta,\bta)=\det D_1\,e^{\bta D_2^{-1}\eta}
\label{2}
\ee
Writing this as 
\be
Z_{f,mixed}(\eta,\bta)=e^{\Delta S}\,\det D_2\,e^{\bta D_2^{-1}\eta}
\label{3}
\ee
where 
\be
\Delta S=Tr\log D_1-Tr\log D_2
\label{4}
\ee
we see that the full lattice QCD theory with mixed fermion action is equivalent
to the lattice fermion being described solely by $D_2$ and the lattice gauge
field action being shifted by
\be
S_{gauge}\;\to\;S_{gauge}+\Delta S\,.
\label{5}
\ee
If the shift (\ref{5}) does not change the universality class, i.e., leaves the
lattice theory in the right universality class for QCD, then surely the same
is true for the smaller shift
\be
S_{gauge}\;\to\;S_{gauge}+{\textstyle \frac{1}{2}}\Delta S\,.
\label{6}
\ee
But this shift is equivalent to leaving $S_{gauge}$ unchanged and changing the
fermion determinant in the mixed fermion generating functional (\ref{2}) by
\be
\det D_1\;\to\;(\det D_1 \det D_2)^{1/2}
\label{7}
\ee
We conclude that if the lattice QCD theory with mixed fermion action is in
the right universality class for QCD then so is the theory where a dynamical
quark is described by the 2nd taste (flavor) of the 2-taste lattice Dirac
operator 
\be
D=\bigg({D_1 \atop 0}\;{0 \atop D_2}\bigg)
\label{8}
\ee
and with rooted fermion determinant $(\det D)^{1/2}$.

Normally there would be no reason to consider such a formulation in practice
rather then just using $D_1$ or $D_2$ or the mixed fermion formulation.
But it is useful to consider this formulation for theoretical investigation
of lattice QCD with rooted fermion determinants -- it is simpler than the
relevant case of staggered fermions since the taste (flavor) interpretation is 
manifest from the beginning and there is no mixing between the 
different tastes. In the next section we will exhibit a 2-taste lattice Dirac
operator of the form (\ref{8}) with properties analogous to the staggered
Dirac operator and for which Creutz's objections also apply. The preceding 
considerations have already shown (at least heuristically) that the viability 
of using the rooted determinant of such an operator is assured if the related 
unrooted mixed fermion theory is in the right universality class for QCD.

\section{A 2-taste lattice Dirac operator with exact chiral symmetries}

The specific 2-taste lattice Dirac operator we will study is given in
the massless case by   
\be
D=\bigg({D_+ \atop 0}\;{0 \atop D_-}\bigg)\quad,\qquad
D_{\pm}=\gmu\nabla_{\mu}\pm i\g5(a{\textstyle \frac{1}{2}}\Delta+m_5)
\label{9}
\ee
where $a=$lattice spacing, $\nabla_{\mu}$ is the usual symmetrized lattice
covariant derivative and $\Delta$ the usual lattice Laplace operator. 
For reasons discussed below we have included a mass parameter $m_5$ in the
operator. It should not be confused with the usual mass; we introduce the 
latter in the usual way: the massive 2-taste Dirac operator is $D+m$. 
Note that $D_{\pm}\,$, and hence $D$, are anti-hermitian, and that in the 
free field case (link variables set to unity) 
\be
\Big\lb\,D_{\pm}^{\dagger}D_{\pm}\,\Big\rb_{free}
=\Big\lb\,\Sn^{\dagger}\Sn+(a{\textstyle \frac{1}{2}}\Delta+m_5)^2
\,\Big\rb_{free}
\label{10}
\ee
which shows that $D_{\pm}$ is free of fermion doubling so $D$ describes two
lattice fermion tastes as claimed. Writing $D$ as
\be
D=(\gmu\otimes{\bf 1})\nabla_{\mu}
+i(\g5\otimes\sigma_3)(a{\textstyle \frac{1}{2}}\Delta+m_5)
\label{11}
\ee
we see that it has the taste-nonsinglet chiral symmetries
\be
\lbrace D,\Gamma_j\rbrace=0\quad,\quad\ \Gamma_j=\g5\otimes\sigma_j\ ,\ j=1,2
\label{12}
\ee
where $\sigma_1$, $\sigma_2$, $\sigma_3$ are the Pauli matrices acting on
taste space. On the other hand, $D$ breaks the symmetry of the chiral 
transformation generated by $\g5\otimes\sigma_3$ (and also $\g5\otimes{\bf 1}$
as it should to produce the axial anomaly). Consequently the pion spectrum
will not be SU(2)-symmetric, so the fermion theory described by $D$ is not 
equivalent at non-zero lattice spacing
to a two-flavor theory with both flavors described by the same 
single-taste lattice Dirac operator.\footnote{This is the same reasoning
that was used in Ref.\cite{BGS} to draw the analogous conclusion for the 
staggered fermion theory.}

Thus $D$ shares key properties with the massless staggered lattice
Dirac operator: it is anti-hermitian and has taste-nonsinglet chiral symmetries
which protect against additive mass renormalization and are expected to be
spontaneously broken just like the taste-nonsinglet U(1) chiral symmetry of
staggered fermions, while other chiral symmetries are explicitly broken at
non-vanishing lattice spacing.
In fact the expression (\ref{11}) has clear similarities
with the massless staggered Dirac operator in the flavor (taste) 
representation \cite{G,KS}. 

However, there is also a significant difference: Our operator breaks the parity 
and time reversal symmetries, since the ``Wilson-like'' term in $D_{\pm}$ 
gives a pseudo-scalar term in the fermion action. Consequently, 
radiative corrections will generate a pseudoscalar mass term. Therefore we 
have included a bare pseudo-scalar mass term with mass $m_5$ in 
(\ref{9}); it should be
tuned to a critical negative value as the continuum limit is approached so as
to cancel the pseudo-scalar mass term generated by radiative corrections and
thereby restore the $P$ and $T$ symmetries in the continuum limit. (This is
analogous to the tuning of the bare scalar mass to a critical negative value 
to reach the chiral limit with usual Wilson fermions.) Through this we also
expect the chiral transformation generated by $\g5\otimes\sigma_3$ to 
become a symmetry of the 2-taste theory in the continuum limit. Usually we will
suppress the $m_5$-dependence of $D$ in the notation, although sometimes we will 
indicate it explicitly as $D(m_5)$.

The 2-flavor theory described by $D(m_5)+m$ can be obtained from the 2-flavor
Wilson fermion theory with twisted mass \cite{twisted} by a flavored chiral 
rotation of the fields. We show this in Appendix A. 
However, as shown there, the symmetries of the theories have different 
interpretations: the chiral symmetries (\ref{12}) correspond to vector
symmetries in the Wilson case.

A major advantage that our setting has over the staggered one for 
investigating the viability of rooting is that there
is a single-flavor fermion theory that our rooted theory can be explicitly 
compared with, namely the the theory described by $D_++m$ (or $D_-+m$).
Comparison of the rooted theory based on $D+m$ with the single-flavor theory 
described by $D_++m$ will be our main focus in this paper. Through this we will
be able to counter Creutz's objections to the rooting trick quite explicitly.

The starting point for much of Creutz's argumentation against the rooted
staggered fermion determinant is the observation that, as a consequence of
its exact $U(1)$ chiral symmetry, the staggered fermion theory with mass $m$ 
is equivalent to the one with mass term changed by
\be
m\bpsi\psi\;\to\;m\bpsi\psi\cos(2\theta)+im\bpsi\Gamma_5\psi\sin(2\theta)
\label{14}
\ee
for any $\theta$, where $\Gamma_5$ is the operator on staggered fermion fields 
corresponding to $\g5$ in the naive lattice fermion theory from which the
staggered theory originates. See Eq.(4) of Ref.\cite{Creutz(chiral)}. 
In particular, $m$, $-m$ and $\pm i\Gamma_5m$ are all physically equivalent.
Therefore, the rooted staggered fermion determinant is invariant under these
changes in the mass term, unlike the determinant of a genuine one-flavor
lattice Dirac operator. Exactly the same is true in the present 2-taste
theory when $\Gamma_5$ in (\ref{14}) is replaced by our $\Gamma_1$ or $\Gamma_2$
in light of (\ref{12}). Therefore, all Creutz's objections against rooting based 
on (\ref{14}) apply just as much in our case. 
In the following sections we derive explicit relations between the rooted 
determinant formulation and single-taste theories in the present case which
show that, despite Creutz's concerns, the rooted formulation does 
appear to be viable, or at least a good approximation when the quark masses are
not too small.

\section{Properties of the single-taste theory and implications for the 
rooted formulation}

In this section we derive general properties of the single-taste theories
described by $D_{\pm}+m$ and use them to obtain significant indications of the
viability of the rooted determinant $\det(D+m)^{1/2}\,$. Throughout the following
we assume that the lattice is finite; then the vectorspace of lattice spinor
fields is finite-dimensional and the fermion determinants are all finite.

\subsection{Fermion determinants and would-be zero-modes}

$D_{\pm}$ in (\ref{9}) satisfies 
\be
D_{\pm}\g5=-\g5 D_{\mp}
\label{15}
\ee
implying an equivalence between the eigenvalue equations for $D_+$ and $D_-\,$:
\be 
D_+\psi_{\lambda}=i\lambda\psi_{\lambda}
\;\Leftrightarrow\;D_-(\g5\psi_{\lambda})=-i\lambda(\g5\psi_{\lambda})
\label{16}
\ee
Using this, we find that the rooted determinant of $D+m$ is given in terms 
of the eigenvalues $\{i\lambda\}$ of $D_+$ by
\be
\det(D+m)^{1/2}=\prod_{\lambda}\sqrt{\lambda^2+m^2}
\label{17}
\ee
Comparing this with the determinant of the single-taste Dirac operator $D_+$
\be
\det(D_++m)=\prod_{\lambda}(i\lambda+m)
\label{18}
\ee
we see that
\be
\det(D+m)^{1/2}=|\det(D_++m)|
\label{19}
\ee
This shows that using the rooted determinant is the same as removing the complex 
phase of the single-taste determinant $\det(D_++m)$. We will show further below
that the single-taste determinant is indeed complex-valued, calculate its
complex phase when $m$ is in a ``chiral region'', 
and discuss how the complex phase can be removed to arrive at the rooted 
determinant via (\ref{19}).
The would-be zero-modes of the 1-taste and 2-taste theories play an important 
role in this, and we begin by considering them in the following.

An exact zero-mode would give rise to a factor $|m|$ in (\ref{19}) rather than
the factor $m$ which would appear in a genuine one-flavor fermion determinant.
This difference is expected (see, e.g., \cite{Sharpe(06),Kronfeld(07)}) and is 
inconsequential as long as considerations are restricted to positive $m$. 
In practice though we do not expect exact zero-modes for this operator; the most
one can hope for is approximate, would-be zero-modes that become exact in the 
continuum limit. To produce expected non-perturbative effects it is crucial
that there are robust would-be zero-modes in topologically non-trivial 
gauge field backgrounds in accordance with the Index Theorem. These are indeed
present in this case, as we will now show.

It is useful to introduce the new gamma-matrices  
\be 
\tg_{\mu}=-i\g5\gmu
\label{20}
\ee
These form another representation of the Dirac algebra: 
$\{\tg_{\mu}\,,\tg_{\nu}\}=2\delta_{\mu\nu}$, continue to be hermitian
($\tg_{\mu}^{\dagger}=\tg_{\mu}$) and have the same chirality matrix as before:
$\tg_5=\g5$. Then $D_+$ in (\ref{9}) can be expressed as 
\be
D_+&=&i\g5(\tg_{\mu}\nabla_{\mu}+{\textstyle \frac{a}{2}}\Delta+m_5) \nonumber \\
&=&i\tg_5(D_{\tW}+m_5)
\label{21}
\ee
where $D_{\tW}$ is the massless Wilson-Dirac operator constructed with the   
new gamma-matrices.
Introducing the hermitian operator
\be
H(m)=\tg_5(D_{\tW}-m)
\label{22}
\ee
we have 
\be
D_+=iH(-m_5)
\label{23}
\ee
so the solutions to the eigenvalue equation
\be
H(m)\psi_{\lambda}(m)=\lambda(m)\psi_{\lambda}(m)
\label{24}
\ee
give back the eigenvalues and eigenvectors of $D_+$ in (\ref{16}) as a special 
case: $\lambda=\lambda(-m_5)\,$, $\psi_{\lambda}=\psi_{\lambda}(-m_5)$. 

From (\ref{22}) we see that
\be
\lambda(m_0)=0\;\Leftrightarrow\;
D_{\tW}\psi_{\lambda}(m_0)=m_0\psi_{\lambda}(m_0)
\label{25}
\ee
i.e., vanishing of an eigenvalue $\lambda(m)$ at $m_0$ corresponds to a real
eigenvalue $m_0$ of the Wilson-Dirac operator with eigenvector 
$\psi_{\lambda}(m_0)$. It is well-known that the would-be zero-modes of the 
Wilson-Dirac operator are precisely the low-lying real (necessarily positive)
eigenvalue modes. As (\ref{25}) shows, these correspond to crossings of 
the origin close to zero (i.e., at some small positive value $m_0$) by 
eigenvalues $\lambda(m)$ of $H(m)$. Furthermore, the low-lying real modes
of $D_{\tW}$ have approximate $\pm$ chirality under $\tg_5\,$, and from (\ref{22})
it is clear that the sign of the chirality is minus the sign of the
slope of $\lambda(m)$ where it crosses the origin at $m_0$. This is all 
well-known and was discussed a long time ago by Itoh, Iwasaki and Yoshi\'e 
\cite{Itoh}. It allows a robust, integer-valued index to be defined for the
Wilson-Dirac operator in terms of the spectral flow of $H(m)$ in the small $m$
region: it is the difference between the number of negative and positive slope 
eigenvalue crossings. In fact this coincides with the index of the overlap
Dirac operator \cite{Neu(overlap)}. It has been studied numerically in
\cite{Itoh,Heller-Narayanan}, and analytically in 
\cite{DA(AnnPhys),DA(JMP)} where it
was shown to coincide with the topological charge of the (smooth) lattice
gauge field in the continuum limit in accordance with the 
Index Theorem.\footnote{The robustness of low-lying real eigenvalue modes and 
index of the Wilson-Dirac operator is ensured by the property that in sufficiently
smooth backgrounds the eigenvalues cannot vary arbitrarily under deformations of 
the background but are constrained to be close to zero. An upper bound can be 
analytically derived when the plaquette variables satisfy a bound 
$||1-U_{\mu\nu}-1||<\e$ \cite{L(local),Neu(bound)}. More generally, a bound
constraining the real eigenvalues to lie in neighborhoods
of $0,\,2/a,\,4/a,\,6/a,\,8/a$ can be derived in this case \cite{DA(bound)}.}

By (\ref{23})--(\ref{25}) a real mode of $D_{\tW}$ with eigenvalue $m_0$
is an exact zero-mode of $D_+(m_5)$ when we set $m_5=-m_0$. For an ensemble
of lattice gauge fields generated at sufficiently small bare coupling (or
with a sufficiently improved lattice gauge action) the low-lying real
eigenvalues of $D_{\tW}$ cluster around a critical (positive) value $m_c$ -- see, 
e.g., \cite{Itoh}. We henceforth tune $m_5$ so that 
\be
m_5=-m_c\,. 
\label{26}
\ee
Consequently the eigenvalues and eigenvectors of $D_+$ are $i\lambda(m_c)$ and
$\psi_{\lambda}(m_c)$, respectively.
Then the would-be chiral zero-modes of $D_{\tW}$ are in one-to-one correspondence
with would-be chiral zero-modes of $D_+$. This is seen as follows.
A low-lying real mode of $D_{\tW}$ with eigenvalue $m_0$ is a zero-mode 
$\psi_{\lambda}(m_0)$ for $H(m_0)$ with approximately definite chirality. 
Since $\lambda(m_0)=0$ and $m_0$ is very close to $m_c$ it follows that
$\lambda(m_c)$ is very small, and the corresponding eigenvector $\psi(m_c)$
is very close to $\psi(m_0)$ and therefore has the same approximate chirality.  
Recall that the sign of the chirality is the opposite of the sign of the slope
of $\lambda(m)$ at $m=m_0$. Therefore, $\pm$ chirality corresponds to 
the sign of $\lambda(m_c)$ being $\pm$ if $m_c<m_0$ and $\mp$ if $m_c>m_0$.

Thus, with $m_5$ tuned as dictated by (\ref{26}), the low-lying modes of 
$D_+$ are generically would-be chiral zero-modes; they are robust since they
are tied to the would-be chiral zero-modes (i.e., the low-lying real modes) of 
the Wilson-Dirac operator $D_{\tW}$. By (\ref{16}) the same is true for 
$D_-$. Note that a would-be chiral zero-mode for the Wilson-Dirac operator 
corresponds to would-be chiral zero-modes for each of 
$D_+$ and $D_-$ with the {\em same} chirality;
consequently it corresponds to two would-be chiral zero-modes for $D$ with the
same chirality. Thus we have established that $D_+\,$, $D_-$, and $D$ all
have robust would-be chiral zero-modes in sufficiently smooth gauge backgrounds,
and that the index defined from these equals the topological charge $Q$ of
the gauge background (or $2Q$ in the case of the 2-taste operator $D$) in
accordance with the Index Theorem, since this holds for the Wilson-Dirac operator.

The tuning of $m_5$ dictated by (\ref{26}) is also the appropriate one for
restoring in the continuum limit the $P$ and $T$ symmetries and the chiral 
symmetry of the 2-taste theory generated by $\g5\otimes\sigma_3$. Restoring
these symmetries means tuning $m_5$ so that the effective pseudo-scalar
mass term vanishes. The usual signal for the vanishing of an effective mass term
(scalar or pseudo-scalar) is divergence of the propagator, and from the
discussion leading to (\ref{26}) it is clear that this tuning produces the
desired divergence in the present case. (It is analogous 
to the tuning of the bare mass to a critical negative value to to reach the 
chiral limit with Wilson fermions.)

Returning now to the fermion determinants, we first note that in approaching 
the chiral limit the scalar mass $m$ should be tuned such that
\be
|\lambda_{low}|\;<<\;|m|\;<<\;|\lambda_{nonlow}|
\label{27}
\ee
where $\lambda_{low}$ refers to the eigenvalues of the would-be zero-modes
and $\lambda_{nonlow}$ refers to all the other eigenvalues. The reason is to
achieve appropriate near-chiral limit mass dependence
in the fermion determinants: from (\ref{17})--(\ref{18}) we see that 
(\ref{27}) is necessary and sufficient to get
\be
\det(D_++m)&\approx&\prod_{\lambda_{low}}m\prod_{\lambda_{nonlow}}
i\lambda_{nonlow} \label{28} \\
\det(D+m)^{1/2}&\approx&\prod_{\lambda_{low}}|m|\prod_{\lambda_{nonlow}}
|\lambda_{nonlow}|
\label{29}
\ee
To compare the rooted determinant with the single-taste determinant
$\det(D_++m)$ we need to determine the complex phase of the latter. We now
calculate it for $m$ in the chiral region (\ref{27}), starting from
\be
\frac{\det(D_++m)}{|\det(D_++m)|}
%&=&\prod_{\lambda}\frac{i\lambda+m}{\sqrt{\lambda^2+m^2}} \nonumber \\
\;\approx\;\prod_{\lambda_{low}}\frac{m}{|m|}
\prod_{\lambda_{nonlow}}i\frac{\lambda_{nonlow}}{|\lambda_{nonlow}|}
%\ +\ O({\textstyle \frac{\lambda_{low}}{m}}\,,\,
%{\textstyle \frac{m}{\lambda_{nonlow}}})
\label{30}
\ee
where negligible terms $\sim\frac{\lambda_{low}}{m}$ and 
$\sim\frac{m}{\lambda_{nonlow}}$ have been dropped.
To evaluate this we will use
\be
\prod_{\lambda_{nonlow}}i=\prod_{\lambda_{low}}(-i)
\label{31}
\ee
which follows from $\prod_{\lambda}i=1$, a consequence of the fact that the 
dimension of the vectorspace of lattice spinor fields is a multiple of 4.
Now recall that the eigenvalues $\lambda_{nonlow}$ are the values at $m=m_c$ of 
$\lambda_{nonlow}(m)$. Generically these do not cross zero in the small $m$ 
region; in particular they do not cross zero as $m$ varies from 0 to $m_c$. 
Therefore,
\be   
\prod_{\lambda_{nonlow}}\frac{\lambda_{nonlow}(m_c)}{|\lambda_{nonlow}(m_c)|}
=\prod_{\lambda_{nonlow}}\frac{\lambda_{nonlow}(0)}{|\lambda_{nonlow}(0)|}
=\prod_{\lambda_{low}}\frac{\lambda_{low}(0)}{|\lambda_{low}(0)|}
\label{32}
\ee
where the last equality follows from 
$\prod_{\lambda}\frac{\lambda(0)}{|\lambda(0)|}=1$, a consequence of the known
fact that $Tr\frac{H(0)}{|H(0)|}=0$; see, e.g., \cite{DA(bound)}.
On the other hand, the eigenvalues $\lambda_{low}(m)$ do cross zero in the small
positive $m$ region. If $\lambda_{low}(0)>0$ then $\lambda_{low}(m)$ has 
negative crossing slope, corresponding to a positive chirality would-be zero-mode
by our previous discussion. Similarly, $\lambda_{low}(0)<0$ implies a would-be
zero-mode with negative chirality. It follows that
\be
\prod_{\lambda_{low}}\frac{\lambda_{low}(0)}{|\lambda_{low}(0)|}=(-1)^{n_-}
\label{33}
\ee  
where $n_{\pm}$ denotes the number of $\pm$ chirality would-be zero-modes.
This together with (\ref{32}) and (\ref{31}) leads to
\be
\prod_{\lambda_{nonlow}}\,i\frac{\lambda_{nonlow}}{|\lambda_{nonlow}|}
=i^{-Q}
\label{34}
\ee
where 
\be
Q=n_+-n_-
\label{35}
\ee
is the index of the would-be chiral zero-modes and coincides with the topological
charge in sufficiently smooth gauge backgrounds as discussed earlier.
Using this in (\ref{30}) we finally obtain
\be
\frac{\det(D_++m)}{|\det(D_++m)|}
\;\approx\;i^{-\frac{m}{|m|}Q}=
e^{-i\frac{m}{|m|}\frac{\pi}{2}Q}
\label{37}
\ee
The equality becomes exact in the limit $\frac{\lambda_{low}}{m}\to0\,$,
$\frac{m}{\lambda_{nonlow}}\to0$, which should be regarded as the chiral 
limit in this setting. The prospects for the possibility of being able to
take this limit (in principle) are discussed in the concluding section.

Thus for $m$ in the chiral region (\ref{27}) the effect of the complex phase of 
$\det(D_++m)$ is to shift the QCD theta-vacuum angle by 
$\theta\to\theta-\frac{m}{|m|}\frac{\pi}{2}$. 
Since the physical theta-vacuum angle must be zero (or extremely close to
zero) \cite{textbook}, the bare theta-vacuum angle in the lattice QCD theory must 
be chosen such that the shifted one vanishes. This is equivalent to having 
trivial theta vacuum and replacing the fermion determinant (with $m>0$) by
\be
\det(D_++m)\;\to\;e^{i\frac{\pi}{2}Q}\,\det(D_++m)
\label{38}
\ee
which is essentially the same as
\be
\det(D_++m)\;\to\;|\det(D_++m)|=\det(D+m)^{1/2}
\label{39}
\ee
in the chiral region.
This strongly indicates the viability of using the rooted determinant to 
represent the determinant for a single quark flavor in the present case.

As a further indication of the viability of the rooted determinant we see from 
(\ref{21}) with (\ref{26}) that
\be
\det D_+=\det(D_W-m_c)\,.
\label{40}
\ee
Since the Wilson fermion determinant is real and positive, this shows that
$\det(D+m)^{1/2}$ coincides at $m=0$ with the Wilson fermion determinant with
bare mass tuned to precisely the negative critical value which it should have
in the chiral limit. Therefore, for very small $m$ in the chiral region 
(\ref{27}), the rooted determinant is very close to the chiral limit of the
Wilson fermion determinant.

Flipping the sign of $m$ has the effect of complex conjugation on the phase 
factor in (\ref{37}). We note in passing that this is a general property
the single-taste fermion determinant: From (\ref{18}) and (\ref{16})
we easily find
\be
\det(D_+-m)=\det(D_++m)^*=\det(D_-+m)
\label{40a}
\ee

\subsection{Origin of the complex phase}

For the rooted formulation to be viable, the low energy physics it describes 
should be the same as when the fermion is described by $D_++m$ with a bare theta
term included in the lattice QCD action to cancel the one produced by 
$\det(D_++m)$. For this to hold, the complex phase should originate from the 
ultra-violet part of the spectrum of $D_++m$, so that it is not a manifestation of 
low energy aspects of the fermion formulation described by $D_++m$. We show this to 
be the case in the following.

%For the rooted determinant $\det(D+m)^{1/2}$ to be a viable representation of
%a single quark determinant the low energy physics of the two tastes described by
%$D+m$ should be physically equivalent in the continuum limit. The tastes are 
%described by $D_++m$ and $D_-+m$ respectively. We have already found significant
%evidence for their equivalence: In a given (sufficiently smooth) gauge background
%they have the same number of would-be zero-modes with the same chiralities in
%accordance with the Index Theorem. Also, they both reproduce the correct axial 
%anomaly in the $a\to0$ limit, as discussed in the next subsection. 
%However, their fermion determinants have complex
%phases which differ, being related to each other through complex conjugation
%cf.~(\ref{40a}). It follows from (\ref{37}) that 
%\be
%\det(D_{\pm}+m)\;\approx\;e^{\mp i\frac{m}{|m|}\frac{\pi}{2}Q}\det(D+m)^{1/2}
%\label{5.1}
%\ee
%for $m$ in the chiral region.
%To check the physical equivalence of the two tastes it is important to find
%the origin of the complex phase and the phase difference between the $+$ and $-$ 
%cases. If it originates from a physically relevant part of the spectrum
%then equivalence of the tastes cannot be asserted. However, 
%this turns out not to be the case, as we now show.

In sufficiently smooth gauge backgrounds where the Index Theorem relation between 
chirality of would-be zero-modes and topological charge holds, it is known
that for each would-be chiral zero-mode of the Wilson-Dirac operator there
are 15 ``doubler'' modes \cite{Itoh,DA(bound)}. 
These are eigenvectors of $D_W$ with approximately 
definite chirality and with large (positive) real eigenvalues clustered around
specific values: If the approximate chirality of the zero-mode is $\pm$
then the associated real eigenmodes consist of four eigenvectors with eigenvalues
$\approx 2/a$ and chirality $\mp$; six eigenvectors with eigenvalues
$\approx 4/a$ and chirality $\pm$; four eigenvectors with eigenvalues 
$\approx 6/a$ and chirality $\mp$, and one eigenvector with eigenvalue 
$\approx 8/a$ and chirality $\pm$ \cite{DA(bound)}. 
By (\ref{25}) this implies a corresponding family 
$\{\psi_j(m)\}_{j=0,1,\dots,15}$ of eigenvectors, 
$H(m)\psi_j(m)=\lambda_j(m)\psi_j(m)$, with each $\lambda_j(m)$ vanishing  
at a value $m_j$ close to $2p/a$ for some $p\in\{0,1,2,3,4\}$. There are 
$\frac{4!}{p!(4-p)!}$ $j$'s for each $p$, and $\psi_j(m_j)$ has the approximate
chirality $\pm(-1)^p$ under $\g5$.

Noting that $H(m_c)$ can be written for each $j$ as $H(m_c)=H(m_j)+(m_j-m_c)\g5$
we see that each $\psi_j(m_j)$ is an approximate eigenvector for $H(m_c)$,
and therefor also for $D_+=iH(m_c)$:
\be
H(m_c)\psi_j(m_j)\;\approx\;\pm(-1)^p(m_j-m_c)\psi_j(m_j)\,.
\label{5.2}
\ee
Since $m_c\approx0$ it follows that, generically, the spectrum $\{i\lambda\}$
of $D_+$ contains 15 ``doubler'' eigenvalues associated with each would-be
zero-eigenvalue; four of them are $\approx\mp i2/a$; six of them are 
$\approx\pm i4/a$; four of them are $\approx\mp i6/a$; and the final one is 
$\approx\pm i8/a$ where $\pm$ is the chirality of the would-be zero-mode.
The contribution of these to the phase factor in (\ref{30}) is
\be
\prod_{j=1}^{15}i\frac{\lambda_j}{|\lambda_j|}
=(\mp i)^4(\pm i)^6(\mp i)^4(\pm i)=\mp i
\label{5.3}
\ee
It follows that the total contribution to the phase factor from the doubler modes 
of all the would-be zero-modes is $(-i)^{n_+}i^{n_-}=i^{-Q}$. This reproduces
precisely the phase factor in (\ref{34}), which for $m>0$ gives the complex
phase of the determinant in (\ref{37}). Thus we have found that, at least when
$m$ is in the chiral region, the complex phase of $\det(D_++m)$ originates
entirely from the would-be doubler modes associated with the would-be
zero-modes of $D_+$. An analogous result holds for $\det(D_-+m)$.
The doubler mode eigenvalues are $\sim 1/a$, so we 
conclude that the complex phases of the determinants are not connected with the 
low energy physics of the fermion tastes described by $D_++m$ and $D_-+m$.

\subsection{Determinant phase factor and axial anomaly in the classical 
continuum limit}

A classical continuum limit version of our determinant phase factor result
(\ref{37}) arises as a special case of a previous result of Seiler and 
Stamatescu (SS) \cite{Seiler}. They considered the $m_5=0$ case of the lattice
Dirac operator
\be
D_{\theta}=\gmu\nabla_{\mu}
+e^{i\theta\g5}(a{\textstyle \frac{r}{2}}\Delta+m_5)\,.
\label{41}
\ee
which coincides with our $D_{\pm}$ for $\theta=\pm\pi/2$. SS showed that the 
fermion determinant $\det(D_{\theta}+m)$ produces a theta-vacuum term 
$e^{-i\theta Q}$ in the classical continuum limit (with $m>0$). 
A simple consequence of their specific result, Eq.(19) of \cite{Seiler},  
is\footnote{The minus sign in the exponent of $e^{-i\theta Q}$
is erroneously absent in Eq.(19) of \cite{Seiler}; it should be present
due to minus sign in their Eq.(22).}
\be
\lim_{a\to0}\;\frac{\det(D_{\theta}^A+m)}{|\det(D_{\theta}^A+m)|}
=e^{-i\theta Q}
\label{42}
\ee
where the gauge background is the lattice transcript
of a smooth continuum gauge field $A$ (satisfying certain technical conditions) 
with topological charge $Q$. For $\theta=\pi/2$ this is obviously
a classical continuum limit version of our result (\ref{37}) with $m>0$. 

The result (\ref{42}) is obtained as a straightforward consequence of another 
result of SS, namely that $D_{\theta}+m$ reproduces the correct 
axial anomaly in the classical continuum limit for all values of $\theta$. 
This implies in particular that fermions described by $D_++m$ and $D_-+m$
both reproduce the correct axial anomaly, so the same is true for the 2-flavor
theory described by our $D+m$. 
We emphasize that both tastes reproduce the correct
anomaly with the right sign; they do not have opposite signs and so Creutz's
concern about cancellation of anomalies \cite{Creutz(chiral)} is not realized 
here. Although the considerations of SS were without $m_5\,$, their
results extend almost immediately to $m_5\ne0$. This is because
$m_5$, just like $m$, appear in the final axial anomaly expression through the
dimensionless quantities $am_5$ and $am$ and hence drop out in the $a\to0$
limit. (However, $m$ plays the role of infrared regulator in intermediate
stages of the evaluation and must therefore be non-vanishing and positive.)  
Here $m_5$ and $m$ may either be constant or tuned as a function
of the lattice spacing as long as $am_5\to0\,$, $am\to0$ for $a\to0$.

\section{Pseudoscalar meson propagator in the rooted formulation}

We now consider the pseudoscalar meson propagator, more specifically its
disconnected piece $G^{DC}(x,y)$, which is supposed to solve the $U(1)$ problem 
by being non-vanishing in the chiral limit in topologically nontrivial gauge field
backgrounds and thereby  partially cancelling the connected piece,
resulting in quicker decay, and hence a large mass. This cancellation, which
was already verified a long time ago in the chiral limit with Wilson 
fermions \cite{Itoh}, requires that $G^{DC}(x,y)$ in a fixed topologically
nontrivial gauge background develops a singularity $\sim 1/m^2$
in the chiral limit, produced by the (would-be) zero-modes of the (lattice)
Dirac operator. 

For simplicity we restrict to the one flavor case; then, with our 2-taste $D$
we have
\be
G^{DC}(x,y)={\textstyle \frac{1}{2}}\mbox{tr}\lb(D+m)^{-1}(x,x)
(\g5\otimes{\bf 1})\rb\, 
{\textstyle \frac{1}{2}}\mbox{tr}\lb(D+m)^{-1}(y,y)
(\g5\otimes{\bf 1})\rb
\label{43}
\ee
We have replaced $\mbox{tr}\to\frac{1}{2}\mbox{tr}$ 
compared to the usual expression to take 
account of the two tastes of $D$. It suffices to consider just one of the 
factors $\frac{1}{2}\mbox{tr}[\cdots]$. A simple calculation using 
(\ref{15})-(\ref{16}) gives
\be
{\textstyle \frac{1}{2}}\mbox{tr}\lb(D+m)^{-1}(x,x)
(\g5\otimes{\bf 1})\rb=\sum_{\lambda}\frac{m}{\lambda^2+m^2}\,
\psi_{\lambda}^{\dagger}(x)\g5\psi_{\lambda}(x)
\label{44}
\ee
Exactly the same expression can be (formally) derived in the continuum 
from $\sd\psi_{\lambda}=i\lambda\psi_{\lambda}$ using the fact that
$\sd(\g5\psi_{\lambda})=-i\lambda(\g5\psi_{\lambda})$. However, in the present 
lattice setting we do not have exact zero-modes in general 
so (\ref{44}) and hence $G^{DC}(x,y)$ vanish at $m=0$. 
(This can also be seen directly from the chiral symmetries
in (\ref{12}) since $\Gamma_j$ ($j\!=\!1,2$) commutes with $\g5\otimes{\bf 1}$.
The situation is the same for staggered fermions -- see Sect. VIII.F of 
Ref.\cite{Itoh}.) Clearly the $m\to0$ limit should not be taken before the 
continuum limit here.\footnote{The necessity of taking continuum limit before
chiral limit is well-known for staggered fermions 
\cite{Durr1,Durr2,Bernard(Durr)}.}
The situation is different from Wilson fermions where the chiral limit can be 
reached by tuning the mass to a critical negative value \cite{Itoh}.
In the present case, reaching the chiral limit requires being able to chose $m$ 
in the same way as in our discussion of the fermion determinants in the previous 
section, namely, it should be in the ``chiral region'' (\ref{27}). 
Then (\ref{44}) becomes
\be
{\textstyle \frac{1}{2}}\mbox{tr}\lb(D+m)^{-1}(x,x)
(\g5\otimes{\bf 1})\rb
\;\approx\;\sum_{\lambda_{low}}\frac{1}{m}\,
\psi_{\lambda_{low}}^{\dagger}(x)\g5\psi_{\lambda_{low}}(x)
\label{45}
\ee
which gives the correct chiral limit behavior of (\ref{44}) and hence also 
$G^{DC}(x,y)$. The fact that only the would-be zero-modes contribute in 
(\ref{45}) fits well with the observation from previous numerical
studies that $G^{DC}$ is essentially given by the contribution from 
low-lying modes -- this was seen for staggered fermions in \cite{Kilcup}
and for Wilson fermions using the Hermitian Wilson-Dirac operator in 
\cite{Negele}. 

From (\ref{45}) and (\ref{29}) we see that in the chiral region (\ref{27}) with
positive $m$ the weighted propagator in the rooted theory, 
$\det(D+m)^{1/2}G^{DC}(x,y)$, has the same form and mass dependence 
as obtained from the 't Hooft vertex in the continuum setting. (This is clear,
e.g., from the description of the latter given in \cite{Creutz(thooft)}.) 
This is clearly not the case for values of $m$ which are smaller than specified in 
(\ref{27}) though.

\section{Concluding discussion}

In the rooted fermion formulation based on $D+m$ one would expect that any problem
connected with chirality would show up most clearly in the ``chiral limit''
of small bare mass $m$. We have found no sign of this, having derived quite 
explicit indications of the viability of the rooted formulation when the bare
mass is positive and in the ``chiral region'' (\ref{27}):
\be
|\lambda_{low}|\;<<\;m\;<<\;|\lambda_{nonlow}|
\label{46}
\ee
and in particular in the ``chiral limit''
\be
\frac{\lambda_{low}}{m}\;\to\;0\qquad,\qquad \frac{m}{\lambda_{nonlow}}\;\to\;0
\label{47}
\ee
The existence of the chiral region and limit requires a gap in the eigenvalue 
spectrum of $D$ between
the eigenvalues $\{i\lambda_{low}\}$ of the would-be zero-modes and the
other eigenvalues $\{i\lambda_{nonlow}\}$. It is plausible that such a gap
will open up as the continuum limit (bare coupling $g\to0$) is approached:
In this limit the fluctuations of the low-lying real eigenvalues
of the Wilson-Dirac operator around a critical value $m_c$ should become
smaller and smaller; then the same is true for the fluctuations of 
$\{\lambda_{low}\}$ around zero when $m_5$ is tuned to $-m_c(g)$ (cf.~\S4).
Setting
\be
f_1(g):=\max\lbrace|\lambda_{low}|\rbrace\qquad,\qquad 
f_2(g):=\min\lbrace|\lambda_{nonlow}|\rbrace  
\label{48}
\ee
we expect  
\be
\frac{f_1(g)}{f_2(g)}\;\to\;0 \quad\mbox{for}\ g\to0
\label{49}
\ee
Then, tuning $m$ as a function of the bare coupling by, e.g.,
\be
m(g)=(f_1(g)f_2(g))^{1/2}
\label{50}
\ee
the chiral limit (\ref{47}) is reached as the continuum limit $g\to0$ is taken. 
This implies that a chiral region (\ref{46}) exists for sufficiently small $g$ 
(and also at larger $g$ for highly improved versions of the lattice actions).

While the requirement $m>0$ for the bare mass has been widely recognized
(e.g., in the reviews \cite{Sharpe(06),Kronfeld(07)})\footnote{There is however a 
possibility of extending the rooted formulation with positive $m$ to general 
complex-valued $m$ via the introduction of a theta term, as discussed in the 
staggered fermion case in Ref.~\cite{Durr(complexmass)}.}, we have found here that 
a more stringent condition is required:\footnote{This condition is not at
all surprising, and in fact the necessity of it could already be inferred
in the staggered fermion case from the remarks in sect.~VIII.F of \cite{Itoh}.} 
\be
m>>\mbox{max}\{|\lambda_{low}|\}
\label{51}
\ee
In a lattice formulation of QCD with the fermion determinant for each dynamical 
quark represented by a rooted determinant the dependence
of each bare mass $m_q$ on $g$ is fixed by by renormalization conditions;
e.g., by requiring that the lattice QCD theory gives specified values
for a selection of hadronic mass ratios. In connection with this, 
Creutz has argued \cite{Creutz(upquark)} that the notion of chiral limit
for a single light quark (in practice the up quark) is physically meaningless 
when the other quarks remain massive: He argues that non-perturbative instanton 
effects will produce renormalization scheme-dependent additive corrections to the 
light quark mass. If this is the case then $m_q=0$ is a scheme-dependent 
statement for the bare mass of the light quark. Then there is no physical reason 
why the bare mass must remain positive in a given scheme (i.e., for a given 
choice of renormalization conditions) as the continuum limit is approached, hence
the requirement (\ref{51}) may be violated, in which case
the rooted formulation may fail. On the other hand, if the $u$ and $d$
quarks are taken to have degenerate bare mass then the pion spectrum is
degenerate and the chiral limit is physically well-defined as the limit where
the pions become massless. In this case we can expect to be able to approach this 
limit from within the chiral region (\ref{46}). This applies not only for the 
present formulation (where the product of the degenerate $u$ and $d$ determinants 
are safely represented by the 2-flavor fermion determinant $\det(D+m)$) but also 
for the staggered formulation where the determinant product is represented by the 
square root of the staggered fermion determinant.

Expressions analogous to (\ref{18}) for the rooted determinant and (\ref{44}) 
for $G^{DC}(x,y)$ hold for staggered fermions since the eigenvalues of the
massless staggered Dirac operator come in pairs $\pm i\lambda$. The 
present case is more explicit, since the eigenvalues $\{i\lambda\}$ are 
those of a bona fide single-taste lattice Dirac operator $D_+\,$, whereas no 
such origin is known for the eigenvalues of the staggered Dirac operator.
Nevertheless, the chiral limit issues discussed here are the same for staggered 
fermions. So achieving (\ref{46})--(\ref{47}) in the staggered fermion case is 
also required for taking the chiral limit there. 
It is encouraging with regard to this that numerical studies with improved 
staggered fermions find a clear gap in the spectrum between the 
low-lying would-be zero-eigenvalues and the remainder of the spectrum
\cite{Follana(PRL),Follana(PRD),Durr(spectrum)}.

Having seen in \S4 that the 2-taste lattice Dirac operator $D$ has robust
would-be chiral zero-modes in topologically non-trivial gauge backgrounds
in accordance with the Index Theorem, a natural question is whether the same
is true in the case of staggered fermions. The numerical studies in 
\cite{Follana(PRL),Follana(PRD)} strongly indicate that this is the case. 
In fact, a version of the techniques used in this paper, supplemented with 
further calculations, enables the robust would-be zero-mode result here to also 
be established for staggered fermions, thereby providing a theoretical basis for 
the numerical results of \cite{Follana(PRL),Follana(PRD)}.
This will be presented in a forthcoming paper. 

The 2-flavor fermion formulation specified by the lattice operator $D$ introduced 
here is mathematically equivalent to twisted mass Wilson fermions, but the 
interpretation of the symmetries is different: Two of the flavored vector
symmetries in the Wilson case correspond to the chiral symmetries (\ref{12})
in our case. Other 2-flavor fermion formulations with flavored chiral symmetry 
have recently appeared \cite{Creutz(graphene),Borici}, inspired by graphene 
structure. Their properties were studied 
in \cite{Bedaque} where a general argument was made that 2-flavor (``minimally
doubled'') fermion formulations with an exact chiral symmetry must necessarily 
violate parity or time reversal symmetry. The formulation based on $D$ 
in this paper is another example of this: it has two exact (flavored) chiral 
symmetries and violates $P$ and $T$ symmetry due to a pseudoscalar term in the 
action.

\begin{acknowledgments} 
I thank Prof. Mike Creutz for feedback on the paper,
including reminding me about the work of Seiler and Stamatescu \cite{Seiler}
and mentioning the possibility of a relation to twisted mass Wilson fermions,
and for correspondence on the rooting issue.
I also thank Prof. Steve Sharpe for for feedback, in particular for correcting 
the discussion of the relationship with twisted mass fermions in a previous
version of this paper.
This research is supported by the BK21 program of Seoul National University.
\end{acknowledgments}

\appendix

\section{Relation to twisted mass Wilson fermions}

The twisted mass Wilson formulation for two lattice fermion flavors has the 
action \cite{twisted}
\be
S_{tm}=\bar{\chi}(\gmu\nabla_{\mu}+a{\textstyle \frac{r}{2}}\Delta
+m+i\mu\g5\sigma_3)\chi\,.
\label{a1}
\ee
Noting that 
\be
i\mu\g5\sigma_3=-\mu\,e^{-i\alpha\g5\sigma_3}\quad,\qquad \alpha=\pi/2
\label{a2}
\ee
we see that the flavored chiral rotation of the fields
\be
\chi=e^{i\alpha\g5\sigma_3/2}\,\psi\qquad,\qquad 
\bar{\chi}=\bar{\psi}\,e^{i\alpha\g5\sigma_3/2}
\label{a3}
\ee
leads to
\be
S_{tm}=\bar{\psi}(\gmu\nabla_{\mu}+i\g5\sigma_3(a{\textstyle \frac{r}{2}}
\Delta+m)-\mu)\psi\,. 
\label{a4} 
\ee
This coincides with the action for our 2-flavor theory,
\be
S=\bar{\psi}(D(m_5)+m)\psi
\label{a5}
\ee
with 
\be
m_5\;\to\;m\qquad,\qquad m\;\to\;-\mu
\label{a6}
\ee
At $m=0$ our formulation has the exact flavored chiral symmetries generated by 
$\Gamma_j=\g5\sigma_j\,$ $j\!=\!1,2$ (recall (\ref{12})). 
By (\ref{a4})--(\ref{a6}) $m=0$ in our theory corresponds to $\mu=0$ in the 
twisted mass theory. But from (\ref{a1}) we see that this is just the usual 
2-flavor Wilson theory with mass $m$. Thus the symmetries, which in our 
theory are flavored chiral symmetries, correspond in the twisted mass setting
to non-chiral symmetries of the usual 2-flavor massive Wilson theory with 
vanishing twisted mass. Specifically, these symmetries, which leave 
$\bar{\chi}(D_W+m)\chi$ invariant, are the vector symmetries 
\be
\delta\chi=-i\sigma_j\sigma_3\chi
\qquad,\qquad
\delta\bar{\chi}=\bar{\chi}i\sigma_j\sigma_3
\label{a7}
\ee
for $j=1,2$.


\begin{thebibliography}{XXX}

\bibitem{Davies(PRL)}
C.T.H.~Davies {\em et.~al.}, Phys.~Rev.~Lett.~92:022001, 2004 [hep-lat/0304004];
%%CITATION=HEP-LAT 0304004;%%

\bibitem{MILC}
MILC collaboration: C.~Aubin {\em et al.}, Phys.~Rev.~D 70:114501 (2004)
[hep-lat/0407028]
%%CITATION = HEP-LAT 0407028;%%

\bibitem{Allison}
I.~Allison {\em et.~al.}, Phys.~Rev.~Lett.~94:172001, 2005 [hep-lat/0411027]
%%CITATION = HEP-LAT 0411027;%%

\bibitem{BGS}
C.~Bernard, M.~Golterman and Y.~Shamir, Phys.~Rev.~D 73:114511, 2006
[hep-lat/0604017]
%%CITATION = HEP-LAT 0604017;%%

\bibitem{Shamir}
Y.~Shamir, Phys.~Rev.~D 75:054503, 2007 [hep-lat/0607007]
%%CITATION = HEP-LAT 0607007;%%   

\bibitem{DA(PRD)}
D.H.~Adams, Phys.~Rev.~D 72:114512, 2005 [hep-lat/0411030]
%%CITATION = HEP-LAT 0411030;%%

\bibitem{Peardon}
F.~Maresca and M.~Peardon, hep-lat/0411029
%%CITATION = HEP-LAT 0411029;%% 

\bibitem{DA(04)}
D.H.~Adams, Nucl.~Phys.~Proc.~Suppl.~140 (2005) 148 [hep-lat/0409013]
%%CITATION = HEP-LAT 0409013;%%

\bibitem{Durr(05)}
S.~Durr, PoS LAT2005:021, 2006 [hep-lat/0509026]
%%CITATION = HEP-LAT 0509026;%%

\bibitem{Sharpe(06)}
S.R.~Sharpe, PoS LAT2006:022, 2006 [hep-lat/0610094]
%%CITATION = HEP-LAT 0610094;%%

\bibitem{Kronfeld(07)}
A.S.~Kronfeld, arXiv:0711.0699
%%CITATION = ARXIV:0711.0699;%%

\bibitem{Creutz(07)}
M.~Creutz, arXiv:0708.1295
%%CITATION = ARXIV:0708.1295;%%

\bibitem{thooft}
G.~'t Hooft, Phys.~Rev.~Lett.~37 (1976) 8; 
%%CITATION = PRLTA,37,8;%%
Phys.~Rev.~D 14 (1976) 3432 [Erratum-ibid.~D 18 (1978) 2199]
%%CITATION = PHRVA,D14,3432;%%

\bibitem{Gregory1}
E.B.~Gregory, A.C.~Irving, C.M.~Richards, C.~McNeile and A.~Hart, arXiv:0710.1725
%%CITATION = ARXIV:0710.1725;%%

\bibitem{Gregory2}
E.B.~Gregory, A.C.~Irving, C.M.~Richards and C.~McNeile, arXiv:0709.4224
%%CITATION = ARXIV:0709.4224;%%

\bibitem{Bernard(top)}
C.~Bernard {\em et.al}, arXiv:0710.3124
%%CITATION = ARXIV:0710.3124;%%

\bibitem{Creutz1}
M.~Creutz, hep-lat/0603020
%%CITATION = HEP-LAT 0603020;%%

\bibitem{Creutz2}
M.~Creutz, PoS LAT2006:208, 2006 [hep-lat/0608020]
%%CITATION = HEP-LAT 0608020;%%

\bibitem{Creutz(chiral)}
M.~Creutz, Phys.~Lett.~B 649 (2007) 230 [hep-lat/0701018]
%%CITATION = HEP-LAT 0701018;%%

\bibitem{Creutz(chiral)2nd}
M.~Creutz, {em ibid.} 649 (2007) 241 [arXiv:0704.2016]
%%CITATION = ARXIV:0704.2016;%%

\bibitem{Creutz(thooft)}
M.~Creutz, arXiv:0711.2640
%%CITATION = ARXIV:0711.2640;%%

\bibitem{BGSSrebut1}
C.~Bernard, M.~Golterman, Y.~Shamir and S.R.~Sharpe, Phys.~Lett.~B 
649 (2007) 235 [hep-lat/0603027] 
%%CITATION = HEP-LAT 0603027;%%

\bibitem{BGSSrebut2}
C.~Bernard, M.~Golterman, Y.~Shamir and S.R.~Sharpe, arXiv:0711.0696
%%CITATION = ARVIV:0711.0696;%%

\bibitem{G}
F.~Gliozzi, Nucl.~Phys.~B 204 (1982) 419;
%%CITATION = NUPHA,B204,419;%%

\bibitem{KS}
H.~Kluberg-Stern, A.~Morel, O.~Napoly and B.~Petersson, Nucl.~Phys.~B 
220 (1983) 447
%%CITATION = NUPHA,B220,447;%%

\bibitem{twisted}
R.~Frezzotti, P.A.~Grassi, S.~Sint and P.~Weisz,
JHEP 0108 (2001) 058 [hep-lat/0101001]
%%CITATION = HEP-LAT 0101001;%%

\bibitem{Itoh}
S.~Itoh, Y.~Iwasaki and T.~Yoshi\'e, Phys.~Rev.~D 36 (1987) 527
%%CITATION = PHRVA,D36,527;%%

\bibitem{Neu(overlap)}
H.~Neuberger, Phys.~Lett.~B 417 (1998) 141 [hep-lat/9707022];
%%CITATION = HEP-LAT 9707022;%%

\bibitem{Heller-Narayanan}
R.G.~Edwards, U.M.~Heller and R.~Narayanan, Nucl.~Phys.~B 518 (1998) 319
[hep-lat/9711029]
%%CITATION = HEP-LAT 9711029;%%

\bibitem{DA(AnnPhys)}
D.H.~Adams, Ann.~Phys.~296 (2002) 131 [hep-lat/9812003]
%%CITATION = HEP-LAT 9812003;%%

\bibitem{DA(JMP)}
D.H.~Adams, J.~Math.~Phys.~42 (2001) 5522 [hep-lat/0009026]
%%CITATION = HEP-LAT 0009026;%%

\bibitem{L(local)}
P.~Hernandez, K.~Jansen and M.~L\"uscher, Nucl.~Phys.~B 552 (1999) 363
[hep-lat/9808010]
%%CITATION = HEP-LAT 9808010;%%

\bibitem{Neu(bound)}
H.~Neuberger, Phys.~Rev.~D 61:085015, 2000 [hep-lat/9911004]
%%CITATION = HEP-LAT 9911004;%%

\bibitem{DA(bound)}
D.H.~Adams, Phys.~Rev.~D 68:065009, 2003 [hep-lat/9907005]
%%CITATION=HEP-LAT 9907005;%%

\bibitem{textbook}
For a textbook discussion see, e.g.,
J.F.~Donoghue, E.~Golowich and B.R.~Holstein, ``Dynamics of the Standard Model''
Cambridge University Press, 1992.

\bibitem{Seiler}
E.~Seiler and I.O.~Stamatescu, Phys.~Rev.~D 25 (1982) 2177 
%%CITATION = PHRVA,D25,2177;%%

\bibitem{Durr1}
S.~Durr and C.~Hoelbling, Phys.~Rev.~D 69:034503 (2004) [hep-lat/0311002] 
%%CITATION = HEP-LAT 0311002;%%

\bibitem{Durr2}
S.~Durr and C.~Hoelbling, Phys.~Rev.~D 71:054501 (2005) [hep-lat/0411022] 
%%CITATION = HEP-LAT 0411022;%%

\bibitem{Bernard(Durr)}
C.~Bernard, Phys.~Rev.~D 71:094020 (2005) [hep-lat/0412030]
%%CITATION = HEP-LAT 0412030;%%

\bibitem{Kilcup}
L.~Venkataraman and G.~Kilcup, hep-lat/9711006
%%CITATION = HEP-LAT 9711006;%%

\bibitem{Negele}
H.~Neff, N.~Eicker, Th.~Lippert, J.~Negele and K.~Schilling,
Phys.~Rev.~D 64:114509 (2001) [hep-lat/0106016] 
%%CITATION = HEP-LAT 0106016;%%

\bibitem{Durr(complexmass)}
S.~Durr and C.~Hoelbling, Phys.~Rev.~D 74:014513 (2006) [hep-lat/0604005] 
%%CITATION = HEP-LAT 0604005;%%

\bibitem{Creutz(upquark)}
M.~Creutz, Phys.~Rev.~Lett.~92:162003 (2004) [hep-ph/0312225]
%%CITATION = HEP-PH 0312225;%%

\bibitem{Follana(PRL)}
E.~Follana, A.~Hart and C.T.H.~Davies, Phys.~Rev.~Lett.~93:241601 (2004)
[hep-lat/0406010]
%%CITATION = HEP-LAT 0406010;%%

\bibitem{Follana(PRD)}
E.~Follana, A.~Hart, C.T.H.~Davies, and Q.~Mason, Phys.~Rev.~D 72:054501 (2005)
[hep-lat/0507011]
%%CITATION = HEP-LAT 0507011;%%

\bibitem{Durr(spectrum)}
S.~Durr, C.~Hoelbling and U.~Wenger, Phys.~Rev.~D 70:094502, 2004 
[hep-lat/0406027]
%%CITATION=HEP-LAT 0406027;%%

\bibitem{Creutz(graphene)}
M.~Creutz, JHEP 0804:017 (2008) [arXiv:0712.1201]
%%CITATION = ARXIV:0712.1201;%%

\bibitem{Borici}
A.~Borici, arXiv:0712.4401
%%CITATION = ARXIV:0712.4401;%%

\bibitem{Bedaque}
P.F.~Bedaque, M.I.~Buchoff, B.C.~Tiburzi and A.~Walker-Loud, arXiv:0801.3361
%%CITATION = ARXIV:0801.3361;%%


\end{thebibliography}
\end{document}